\documentclass[epj,amsmath,amssymb]{svjour}
\usepackage{graphicx}

\begin{document}
\newcommand \eq [1]{Eq.~(\ref{#1})}
\newcommand \fig [1]{Fig.~(\ref{#1})}

\title{Weakly driven anomalous diffusion in non-ergodic regime: an analytical solution}

\author{Mauro Bologna\inst{1}\thanks{\email{mauroh69@gmail.com}}
\and Gerardo Aquino\inst{2}
\thanks{\email{g.aquino@imperial.ac.uk}}}

\institute{
  \inst{1} Instituto de Alta Investigaci\'{o}n, Universidad
de Tarapac\'{a}-Casilla 6-D Arica, Chile \\
  \inst{2} Department of
Life Sciences, Imperial College London, SW7 2AZ London, UK}


\date{\today}

\abstract{ We derive the probability density of a diffusion
process generated by nonergodic velocity fluctuations in presence
of a weak potential, using the Liouville equation approach. The
velocity of the diffusing particle undergoes dichotomic
fluctuations with a given distribution~$\psi(\tau)$ of residence
times in each velocity state. We obtain analytical solutions for
the diffusion process in a generic external potential and for a
generic statistics of residence times, including the non-ergodic
regime in which the mean residence time diverges. We show that
these analytical solutions are in agreement with numerical
simulations.}
%

\maketitle

\section{Introduction}
A theoretical study of diffusive transport phenomena has been a
fundamental subject of research since the seminal work on Brownian motion published by Einstein~\cite{einstein}  in 1905,  the first
of four papers of his  {\it annus mirabilis}. The problem was soon
afterwards tackled  by Smoluchowski~\cite{smol} within a more
microscopic picture not based on the assumption of a Stokes-law
for the fluid. Both derivations produce a theoretical description
of ordinary diffusion leading to the well known linear relation
between mean squared displacement and time, $\langle x^2\rangle=Dt$, with~$D$ being the diffusion coefficient. More recently,
several relevant physical and biological conditions have been
discovered where diffusion is characterized by  a non-linear
``anomalous'' relationship between mean squared displacement and
time, $\;\;\;\langle x^2\rangle= D_{\alpha} t^\alpha$. Diffusion through
porous media or within a cellular crowded environment are two
important examples where such anomalous diffusion is detected. In
the past twenty years anomalous diffusion  has therefore become
the subject of intense research work
\cite{strunz,anomalousdiffusion1,anomalousdiffusion2,mioscaf,anomalousdiffusion3}. Herein we aim at developing a theoretical
description for  anomalous diffusion generated by dichotomic
velocity
fluctuations 
based on the following non-linear equation \cite{sancho}

\begin{eqnarray}
 \label{eq1}
\frac{d \mathbf{x}}{dt}=-\texttt{{\boldmath$\nabla$}}
U(\mathbf{x})+ g(\mathbf{x})\texttt{{\boldmath$\xi$}}(t)
\end{eqnarray}
where~$\texttt{{\boldmath$\xi$}}(t)$ can be interpreted as a
stochastic force, that  we shall consider to be a dichotomic
stochastic variable, while~$-\texttt{{\boldmath$\nabla$}}
U(\mathbf{x})$ is the deterministic force. A dependence  $g(\mathbf{x})$ of the amplitude of the stochastic component
is also introduced for  sake of generality, but can be eliminated via a simple change of variables (see following section). We adopt this choice
since the dichotomic process, by taking two well defined finite values
(as opposed to a continuous Gaussian process), is suitable to describe a wide
range of physical processes. Furthermore, it can be shown that, by
an appropriate limit procedure, the dichotomic noise converges to
Gaussian white noise and white shot noise~\cite {van}.

In spite of significant progress reached in the field, the general
solution of \eq{eq1} with an arbitrary potential function
$U(\mathbf{x})$ is still an unsettled problem. Finding such a
solution is the main focus of this paper. In the following we
derive an analytical solution for the probability density function
({\it pdf})~$P(\mathbf{x},t)$, valid for small potential $U(\mathbf{x})$.
 Importantly our solution extends to the case of non-ergodic fluctuations, leading to an elegant description of a regime which is central in  very recent theoretical and experimental explorations, e.g. dynamics of nanocrystals \cite{nature}, dynamics of single molecules in cells \cite{strunz2}, non-markovian stochastic resonance in physical and biological systems \cite{goyc}.
We  confirm all our results by comparison with numerical simulations of the
diffusion process.

\section{Liouville equation}
In this section we consider the one-dimensional version of \eq{eq1} which,
with a simple change of variables $y=h(x)$ with $h(x)=\int^x dz/g(z)$ and consequent redefinition of the potential as 
\begin{equation}
  \tilde{U}(y)=\int^y dx \frac{1}{g(x)}\frac{\partial U}{\partial x},
  \end{equation}
can be conveniently rewritten as
\begin{eqnarray}
 \label{eq2y}
\frac{d y}{dt}=-\frac{\partial }{\partial y}\tilde{U}(y)+
 \xi(t).
\end{eqnarray}
We therefore focus in the rest of the paper on the process with time evolution of general form:
\begin{eqnarray}
 \label{eq2}
\frac{d x}{dt}=-\frac{\partial }{\partial x}U(x)+
 \xi(t).
\end{eqnarray}
and on the corresponding Liouville equation for  the stochastic probability density $\rho(x,t)$, i.e  the probability density for the process 
to get the value $x$ at time $t$ for a given realisation of the fluctuations $\xi(t)$:

\begin{eqnarray}
 \label{eq3}
\frac{\partial }{\partial t}\rho(x,t)=\frac{\partial }{\partial
x}\left[U'(x)-
 \xi(t)\right]\rho(x,t),
\end{eqnarray}
with~$U'(x)\equiv\frac{\partial }{\partial x}U(x)$. 
 Depending on
the sign of~$\xi(t)$  which, without loss of generality, is set to be ~$\xi(t)=\pm 1$, a formal solution of \eq{eq3} is
given by

\begin{eqnarray}
 \label{eq4}
\rho^{(\pm)}(x,t)=\frac{1}{1\mp U'(x)}
 \delta\left(t+\int_0^{x}\frac{1}{U'(z)\mp 1}dz\right).
\end{eqnarray}
 Assuming
that the two values of $\xi$ have the same probability, i.e. 1/2, using
Van Kampen's~\cite{kamp} lemma we can
connect the stochastic density~$\rho$ to the probability density $P(x,t)$ via the relation

\begin{eqnarray}
\nonumber P(x,t)&=&\langle\rho(x,t)\rangle=\frac{1}{2}
\langle\rho^{(+)}(x,t)\rangle+
\frac{1}{2}\langle\rho^{(-)}(x,t)\rangle=
\\
\nonumber &&\frac{1}{1-U'(x)}
 \left\langle\frac{1}{2}
 \delta\left(t+\int_0^{x}\frac{1}{U'(z)- 1}dz\right)\right\rangle+
 \\\label{lemma}&&\frac{1}{1+U'(x)}
 \left\langle
 \frac{1}{2}\delta\left(t+\int_0^{x}\frac{1}{U'(z)+1}dz\right)\right\rangle
\end{eqnarray}
Applying Eq. (\ref{eq4}) to the case of a  linear potential $U(x)=k x$, i.e. a constant force, it follows:

\begin{eqnarray}
 \nonumber 
\rho^{(\pm)}(x,t)&=&\frac{1}{1\mp U'(x)}
 \delta\left(t+\int_0^{x}\frac{1}{U'(z)\mp 1}dz\right)=\\
&& \label{Kforce} \frac{1}{1\mp k}
 \delta\left(t\pm \frac{x}{1\mp k}\right).
\end{eqnarray}
Using the following property of the
Dirac delta function
\begin{equation}
\delta\left[f(z)\right]=\sum_{i}\frac{1}{\mid f'(\bar{z}_i)\mid
}\delta\left(z-\bar{z}_i\right),
\end{equation}
 where~$\bar{z}_i$ are the roots
of the equation~$f(z)=0$,
from Eq. (\ref{Kforce})  we obtain the following exact solution

\begin{eqnarray}
 \label{exactlinear}
\rho^{(\pm)}(x,t)=
 \delta\left[x\pm t(1\pm k)\right]= \delta\left[x+k t\pm t \right]=
 \delta\left[\bar{x}\pm t \right],\;\;\;
\end{eqnarray}
with $\bar{x}=x+kt$.
While in the case of linear potential an exact solution can be found,
in order to find a solution for a generic potential
we consider the case where the deterministic force is a
perturbation, i.e.~$\mid U'(x)\mid\ll 1$. We may rewrite \eq{eq4}
as

\begin{eqnarray}
 \label{eq5}
\rho^{(\pm)}(x,t)\approx\frac{1}{1\mp U'(x)}
 \delta\left(t-U(x)\mp x\right)
\end{eqnarray}
where~$U(x)=\int_0^x U'(z)dz$. 
This approximation remains valid at any time if the force is limited, i.e.~$\mid U'(x)\mid\ll 1$, condition that is necessary to avoid
space confinement which would block  the diffusive transport induced by the stochastic fluctuations $\xi$.
Replacing the variable~$x$ with its functional dependence on $t$,
we may write at the same order in~$U(x)$

\begin{eqnarray}
 \label{eq6}
\rho^{(\pm)}(x,t)\approx
 \delta\left(x\mp t\pm U(\pm t)\right)
\end{eqnarray}
Our main effort is to evaluate the average of the above quantities
over the fluctuations~$\xi$ of the dichotomous random process. Let
us focus on the basic solution for which at~$t=0$ the stochastic
variable is~$\xi=1$

\begin{eqnarray}
 \label{eq7}
\rho^{(+)}_{R}(x,t)=
 \delta\left(x- t+ U(t)\right).
\end{eqnarray}
After a random time~$\tau_1$ generated with a waiting time
distribution~$\psi(\tau)$ the sign of the random variable changes
and we have to consider the second solution, corresponding to
$\xi=-1$, that is to say

\begin{eqnarray}
 \label{eq8}
\rho^{(-)}_{R}(x,t)=
 \delta\left(x+ t- U(-t)+c_1(\tau_1)\right).
\end{eqnarray}
Imposing the continuity condition at the time~$t=\tau_1$ of the two solutions,
(\ref{eq7}) and (\ref{eq8}), we obtain for
the constant~$c_1(\tau_1)$ the expression

\begin{eqnarray}
 \label{eq9}
c_1(\tau_1)=-2\tau_1+U(\tau_1)+U(-\tau_1).
\end{eqnarray}
 In the present paper we shall consider the case
where the potential is an odd function,~$U(z)=-U(-z)$, since this condition leads to cancellation of the potential terms on the righthand side of Eq. (\ref{eq9}),  and therefore to a simple continuity condition of the solutions at each change of $\xi$.  We leave the general
solution for a totally arbitrary potential for
a separate study. From the chosen potential symmetry
it follows that the constant~$c_1$ is

\begin{eqnarray}
 \label{eq10}
c_1(\tau_1)=-2\tau_1.
\end{eqnarray}
Starting with the positive value of the random variable~$\xi(t)=+1$, we have,
 up to time~$\tau_1$, when the random variable
changes sign,

\begin{eqnarray}
 \label{deltas}&&\rho_{R,1}^{(+)}(x,t)=\delta(t-\bar{x})
\theta(t)\theta(\tau_1-t), \\
 \label{deltas2_b}&&\rho_{R,2}^{(-)}(x,t)=
\delta(t+\bar{x}-2\tau_1)\theta(t-\tau_1)
\end{eqnarray}
where, for sake of compactness, we defined 
$\bar{x}=x+U(t)$. Iterating the procedure for the general case  of $n$ changes we
may write the generic terms in the following form
\cite{noi,calisto}

\begin{eqnarray}\nonumber
\rho^{(-)}_{R,2n}(x,t)&=& \delta\left(t+\bar{x}-
2\sum_{k=1}^{2n-1}\tau_k\sin^2\frac{k\pi}{2}\right)\times\\
&&\theta\left(t-\sum_{k=1}^{2n-1}\tau_k\right)\theta
\left(\sum_{k=1}^{2n}\tau_k-t\right), \label{gener}
\end{eqnarray}
and

\begin{eqnarray}\nonumber
\rho^{(+)}_{R,2n+1}(x,t)&=&\delta\left(t-\bar{x}-
2\sum_{k=1}^{2n}\tau_k\cos^2\frac{k\pi}{2}\right)\times\\
&&\theta\left(t-\sum_{i=k}^{2n}\tau_k\right)
\theta\left(\sum_{i=k}^{2n+1}\tau_k-t\right)\label{gener2}.
\end{eqnarray}
Due to the form of the potential, to first order in~$U(x)$,
the solutions coincide with the case of null potential, where we
have to replace the variable~$x$ with~$\bar{x}=x+U(t)$. We now
take the average of the total~$\rho_{R}$ defined as
$\rho_{R}(x,t)=\sum\rho^{(-)}_{R,2n}(x,t)+\sum\rho^{(+)}_{R,2n+1}(x,t)$.
Considering the symmetric case where the waiting time
distribution,~$\psi(\tau)$, for the state~$\xi=1$ is the same
function of the waiting time distribution for the state~$\xi=-1$,
we may perform an average through the multiple integral
$$\int_{0}^{\infty}\prod_{i}
d\tau_i\rho_{R}(x,t)\psi(\tau_{i}).$$ The statistical average
generates the time convoluted expression

\begin{eqnarray}\nonumber
&&\langle\rho_{R}(x,t)=\rangle \frac{1}{2}\left[
\psi_{n}\left(\frac{t-\bar{x}}{2}\right)\times\right.\\\nonumber
&&\left.\int_{0}^{\frac{t+\bar{x}}{2}}
\psi_{n}\left(\frac{t+\bar{x}}{2}-z\right) \Psi(z)dz +
\psi_{n}\left(\frac{t+\bar{x}}{2}\right)\times\right.\\
 &&\left.
\int_{0}^{\frac{t-\bar{x}}{2}}
\psi_{n-1}\left(\frac{t-\bar{x}}{2}-z\right) \Psi(z)dz\right]
\theta(t-\mid\!\bar{x}\!\mid)\label{gener3a},
\end{eqnarray}
where by definition

\begin{eqnarray}
\psi_{0}(z)&\equiv&\delta(z),\\
\psi_{n}(z)&=&\int\limits_{0}^{z}\psi_{n-1}(y)\ast\psi(y)dy,\\
\Psi(z)&=&\int\limits_{z}^{\infty}\psi(y)dy.
\end{eqnarray}
Analogously, for fluctuations starting with ~$\xi=-1$  at~$t=0$, we may write

\begin{eqnarray}\nonumber
&&\langle\rho_{L}(x,t)\rangle=\theta(t-\mid\!\bar{x}\!\mid)\frac{1}{2}\left[
\psi_{n}\left(\frac{t-\bar{x}}{2}\right)\times\right.\\\nonumber &
&\left. \times \int_{0}^{\frac{t+\bar{x}}{2}}
\psi_{n-1}\left(\frac{t+\bar{x}}{2}-z\right)\Psi(z)dz +
\psi_{n}\left(\frac{t+\bar{x}}{2}\right)\times\right.\\
& & \left. \int_{0}^{\frac{t-\bar{x}}{2}}
\psi_{n}\left(\frac{t-\bar{x}}{2}-z\right) \Psi(z)dz\right]
\label{gener3}.
\end{eqnarray}
 The probability density
 function $P(x,t)$, can then be calculated via Eqs.(16) and (\ref{gener3}) and
\begin{equation}
P(x,t)=\frac{1}{2}\langle\rho_{L}(x,t)\rangle+
\frac{1}{2}\langle\rho_{R}(x,t)\rangle\label{gener3b}.
\end{equation}
Summing over the index~$n$ 
and
performing the double Laplace transform, defined as
$$\hat{F}(s,p)=\int_{0}^{\infty}
\int_{0}^{\infty}\exp[-s u]\exp[-vp]F(u,v)dudv,$$ where~$u\equiv
\frac{t-\bar{x}}{2},\,\,v\equiv \frac{t+\bar{x}}{2}$, we find for
the probability density the following expression

\begin{eqnarray}\label{eqc7_b}
\hat{P}(s,u)= \hat{P}_{C}(s,u)+\hat{P}_{D}(s,u).
\end{eqnarray}
Here~$\hat{P}_{C}(s,u)$ refers to the central part of the
distribution~\cite{noi}

\begin{eqnarray}\label{eqc7_bc}
\hat{P}_{C}(s,u)\!=\!\frac{1}{4}\!\left[\!\frac{\hat{\psi}(s)
\left[1\!-\!\hat{\psi}(u)^2\right]}{u [1-\hat{\psi}(s)
\hat{\psi}(u)]}+\frac{\left[1\!-\!\hat{\psi}(s)^2\right]
\hat{\psi}(u)}{s[1-\hat{\psi}(s) \hat{\psi}(u)]}\!\right],
\end{eqnarray}
while~$\hat{P}_{D}(s,u)$ describes the two delta functions

\begin{eqnarray}\label{eqc7_bd}
\hat{P}_{D}(s,u)=\frac{1}{4}\left[ \frac{1-\hat{\psi}(u)}{u}+
\frac{1-\hat{\psi}(s)}{s}\right].
\end{eqnarray}
Eq. (\ref{eqc7_bd}) describes  the particles that did
not change the value of the velocity and in the case of vanishing
potential coincides with the front of particles moving with
uniform velocity.

We note that in general this therm can be evaluated exactly also
in the presence of a potential. Indeed, using the exact expression
given by Eq.~(\ref{eq4}), the first term starting the process,
i.e. Eqs.~(\ref{deltas}) and (\ref{deltas2_b}), can be evaluated
exactly as

\begin{eqnarray}\nonumber
P_{D}(x,t)&=&\frac{1}{2}\Psi(t)\left[ \frac{1}{1- U'(x)}
\delta\left(t+\int_0^{x}\frac{1}{U'(z)- 1}dz\right)
\right.\\\label{deltas2} &+& \left.   \frac{1}{1+ U'(x)}
 \delta\left(t+\int_0^{x}\frac{1}{U'(z)+ 1}dz\right)\right] .
\end{eqnarray}
As pointed out in Ref.~\cite{zum}, we stress that our
approach is based on the assumption that the event occurring at
each random time~$\tau_i$ changes the sign of~$\xi(t)$. We must
therefore refer ourselves to the probability distribution
$\psi(\tau )$. This is the distribution of time intervals between
the changes of value of the variable~$\xi$.

\section{Poissonian velocity fluctuations} \label{tel2}
Here we consider first, the case where the stochastic dichotomous
process generating the velocity fluctuations is exponentially
correlated. A dichotomous renewal process 
with an exponential correlation is characterized by an
an exponential waiting time distribution
density and therefore corresponds to a Poissonian process. The process is
driven by an equation for the probability density known as
telegrapher's equation and it is widely studied including its
generalizations~\cite{metz,ferr}.

To make the paper as self-contained as possible, let us review
briefly hereby the known results based on the correlation
function technique. The correlation function approach has been
successfully used, in particular, for Poisson processes. The~$n$th
correlation function of such a processes fulfills the
condition~\cite{Shapiro}

\begin{eqnarray}\label{shapiro_log}
\frac{\partial}{\partial
t}\langle\xi(t)\xi(t_1)\cdots\xi(t_n)\rangle
=-\gamma\langle\xi(t)\xi(t_1)\cdots\xi(t_n)\rangle
\end{eqnarray}
where the average is performed on the~$\xi$ realizations. The
above formula allows us to write a system of coupled  partial
differential equations for the distribution in the Poisson case.
Averaging Eq.~(\ref{eq3}) over the fluctuations of $\xi$ we may write

\begin{eqnarray}
 \label{eq3bb}
\frac{\partial }{\partial t}\langle\rho(x,t)\rangle=\frac{\partial
}{\partial x}\left[U'(x)\langle\rho(x,t)\rangle-
 \langle\xi(t)\rho(x,t)\rangle\right].
\end{eqnarray}
Introducing the function~$P_1(x,t)=\langle\xi(t)\rho(x,t)\rangle$
and using the differentiation formula for Poisson
processes~\cite{Shapiro}, it follows that

\begin{equation}\label{shap}
\frac{\partial}{\partial t}\langle\xi(t)\phi(t)\rangle=
-\gamma\langle\xi(t)\phi(t)\rangle+\langle\xi(t)\frac{\partial}{\partial
t}\phi(t)\rangle.
\end{equation}
Eqs. (\ref{eq3bb}) and (\ref{shap}) form a system of coupled
partial differential equations with coefficients depending on the
spatial variable $x$

\begin{eqnarray}
 \label{syst1}
&&\!\!\!\!\!\!\!\!\!\!\!\!\!\!\! \partial_t
P(x,t)=\partial_x\left[U'(x)P(x,t)-
 P_1(x,t)\right]\\
 \nonumber  &&\!\!\!\!\!\!\!\!\!\!\!\!\!\!\!\partial_t P_1(x,t)\!=\!-\!\gamma\!
P_1(x,t)\!+\!\partial_x\left[U'(x)\!P_1(x,t)\!-\!P(x,t) \right].
\end{eqnarray}
It is straightforward to obtain the equilibrium distribution for
Eqs.~(\ref{syst1}) 
i.e.

\begin{eqnarray}
 \label{equi}
 P_{eq}(x)=\frac{c}{1-U'(x)^2}\exp\left[- \gamma \int \frac{U'(x)}{1-U'(x)^2}
 dx\right]
\end{eqnarray}
From Eq. (\ref{equi}) we can infer that~$P_{eq}(x)$ is not
always defined and it depends explicitly on the behavior of the
function~$U'(x)$.

The laplace transform with respect to the variable~$t$ of the
coupled system of equations (\ref{syst1})  can
be split in two second order differential equations with variable
coefficients for ~$\hat{P}(s,u)$ and~$\hat{P}_1(s,u)$. It is well
known that  second order differential equations with variable
coefficients can be solved exactly only for a restrict class of
coefficients. The main goal of this section is to show that the
solution given by \eq{eqc7_b} describes  the processes under study
and consequently  provides as well a solution of the system
(\ref{syst1})  for small potential $U$. The
central part of the function~$P(x,t)$, is given by~\cite{noi}

\begin{equation}\label{eqc7b}
\hat{P}_{C}(s,u)=\frac{1}{4} \frac{\gamma\left[2 s u+3 (s+u)
\gamma +4\gamma ^2\right]}{(s+\gamma)(u+\gamma)[s u+(s+u)\gamma]}
\end{equation}
where we used for the waiting time distribution the expression
$\psi(t)=\gamma  \exp\left[-\gamma  t\right]~$. The double inverse
Laplace transform with respect to~$u$ and~$v$ gives the well known
result (see Ref.~\cite{weiss} for a review)

\begin{eqnarray}\nonumber
P_{C}(x,t)&=&\frac{1}{2} \exp\left[-\gamma  t \right] \gamma
\left(I_0\left[\gamma \sqrt{t^2-x^2}\right]+\right.\\\label{eqc8}
&&\left. \frac{t I_1\left[\gamma \sqrt{t^2-x^2}\right]}{
\sqrt{t^2-x^2}}\right)\theta\left(t-\mid\! x\!\mid \right),
\end{eqnarray}
where~$I_n(z)$ is the modified Bessel function of the first kind.
Thus the total \emph{pdf} is:

\begin{eqnarray} \label{eqc9}
P(x,t)=\frac{1}{2}\exp\left[-\gamma t \right]
  \delta\left(t-\mid\!\! x\!\!\mid\right)+
P_{C}(x,t).
\end{eqnarray}
Replacing~$x\to x+U(t)$, the solution for the case with potential
is

\begin{equation}\label{eqc9_b}
P^{(P)}_{U}(x,t)\approx P(x+U(t),t).
\end{equation}
\begin{figure}[ht]
\centering
\includegraphics[ height=5.3 cm, angle=0]{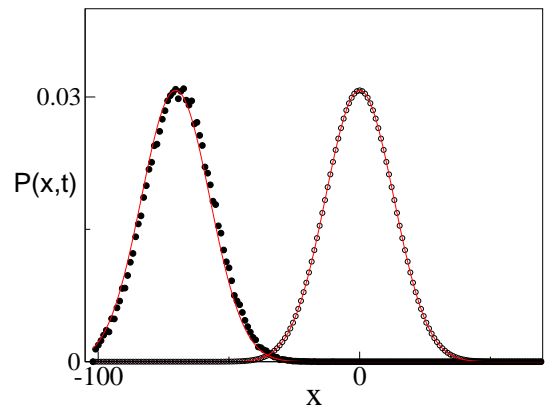}
\caption{Poissonian regime: Density profile for the diffusion process with
fluctuations with exponential waiting times distribution
$\psi(t)=\frac{\gamma}{2} e^{-\gamma t/2}$  with~$\gamma=1$.
Plotted are  the numerical simulation of the diffusion equation
for the case without potential (black empty circles) and  with
potential~$U(x)=\varepsilon\frac{x^3}{1+x^2}$ (black filled
circles) and plots of their respective analytical solution (red
lines) at time~$t=100.5$ with~$\varepsilon=0.1$ (arbitrary units).}
\vspace{0.0 cm}
 \label{fig1a}
\end{figure}

\begin{figure}[ht]
\centering
\includegraphics[height=5.3 cm, angle=0]{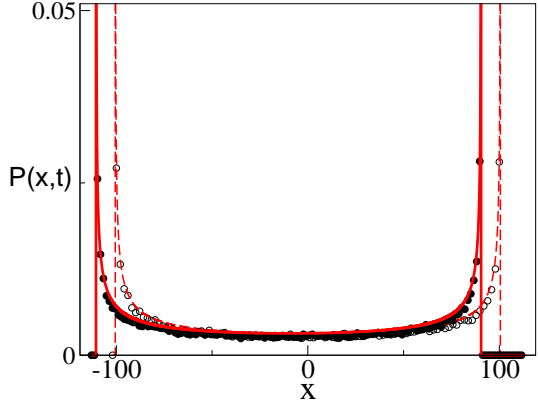}
\caption{Non-ergodic regime: Density profile for the diffusion process with
fluctuations generated with Mittag-Leffler waiting times distribution with
$\alpha=0.5$ and~$T =1$. Plotted are the numerical simulation of
the diffusion equation for the case without potential (black empty
circles) and  with potential~$U(x)=\varepsilon\frac{x^3}{1+x^2}$
(black filled circles) and plots of their respective analytical
solutions (red dashed line and red line) at time~$t=100.5$ with
$\varepsilon=0.1$ (arbitrary units).}\label{fig1}
\end{figure}
We note that the terms containing the two deltas can be evaluated
exactly via \eq{deltas2}.  Also, as an additional result,
~$P_{U}(x,t)$ represents the  solution of the partial differential
equation system (\ref{syst1}) to first order in
the potential $U$. Fig.~1 shows excellent agreement for this
analytical solution with numerical simulations of Eq. (2). It is
straightforward to evaluate the first and the second moment. Using
the definition, for the first moment we have

\begin{equation}\label{fmp}
\langle x\rangle\simeq \int_{-\infty}^{\infty} dx x P(x+U(t),t)\simeq -U(t),
\end{equation}
while for the second moment
\begin{equation}\label{smp}
\langle x^2\rangle\simeq \int_{-\infty}^{\infty}dx x^2
P(x+U(t),t)\simeq \langle x^2\rangle_0 +U(t)^2,
\end{equation}
where $\langle x^2\rangle_0$ is the unperturbed second moment.
 In other words the second moment is affected by the potential
starting from the second order terms in $\varepsilon$, i.e. terms proportional to
$U(t)^2$,  which can be neglected for  limited non-divergent potentials.

\section{Non-ergodic velocity fluctuations} \label{tel3}
We consider here the case of dichotomic fluctuation with waiting
times distributed  with a power-law with index~$\mu<2$. The study
of this condition is of growing interest in the field of out of
equilibrium statistical mechanics, and especially for the linear
response of renewal non-ergodic systems to an external
perturbation \cite{prlbar,prlall,eplprl,eplprl2,prlall2}. A
convenient waiting time distribution for our calculations is given
by the negative derivative of Mittag-Leffler
function~\cite{baet,main}, that is to say
\begin{equation}\label{pleaselabel}
\psi(t)=- \frac{d}{dt}E_{\alpha}
\left[-\left(\frac{t}{T}\right)^{\alpha}\right],\,\,\,\,0<\alpha<1.
\end{equation}
This function has  an asymptotic power-law  behavior, namely
$\psi(t) \sim t^{-\mu}$ for~$t\to\infty$ with~$\mu=\alpha+1$ and its
Laplace transform is

\begin{equation}\label{distr}
\hat{\psi}(s)=\frac{1}{1+(sT)^\alpha}.
\end{equation}
The central part of the distribution has an analytical expression
given by~\cite{lam,noi}

\begin{eqnarray}\nonumber
&&P_C(x,t)=\frac{2\left(1-\frac{x^2}{t^2}\right)^{\alpha-1}\sin\pi
\alpha}{\pi t}\times\\&&\frac{\theta(t-\mid\! x\!\mid )}{
\left(1-\frac{x}{t}\right)^{2 \alpha
}+\left(1+\frac{x}{t}\right)^{2 \alpha }+2
\left(1-\frac{x^2}{t^2}\right)^{\alpha} \cos\pi
\alpha}\label{eqc14}.
\end{eqnarray}
Note that~$P_C(x,t)$ is normalized. Similarly to the Poissonian
case, we may write the expression for the probability density as

\begin{equation}\label{eqc9_bb}
P^{(NP)}_{U}(x,t)\approx P(x+U(t),t).
\end{equation}
Fig.~2 shows comparison of this solution to numerical simulations.
Also we may evaluate the first and the second moment that, as for
the poissonian case, are
\begin{equation}\label{fsmnp}
\langle x\rangle=-U(t),\,\,\,\,\langle x^2\rangle\approx\langle
x^2\rangle_0 +U(t)^2,
\end{equation}
with the quadratic correction in the potential  negligible for limited non-divergent potentials.

\begin{figure}[ht]
\vspace{0.1 cm}
\includegraphics[ height=5.8 cm, angle=0]{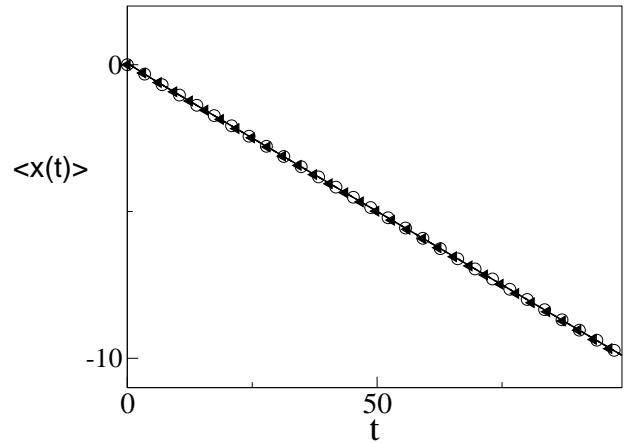}
\caption{ First moment of the distribution as a function of time
in the case of potential~$U(x)=\varepsilon\frac{x^3}{1+x^2}$.
Numerical simulation (dashed and continuous line) vs. 
expression Eqs.~(\ref{fmp}) and ~(\ref{fsmnp}) (filled triangles and
circles) for an exponential waiting times distribution
$\psi(t)=\frac{\gamma}{2} e^{-\gamma t/2}$  ($\gamma=1$) and
Mittag-Leffler distribution ($\alpha=0.5, T =1$) respectively.}
 \label{fig1m}
\end{figure}

\begin{figure}[ht]
\vspace{0.1 cm}
\includegraphics[ height=7.8 cm, angle=270]{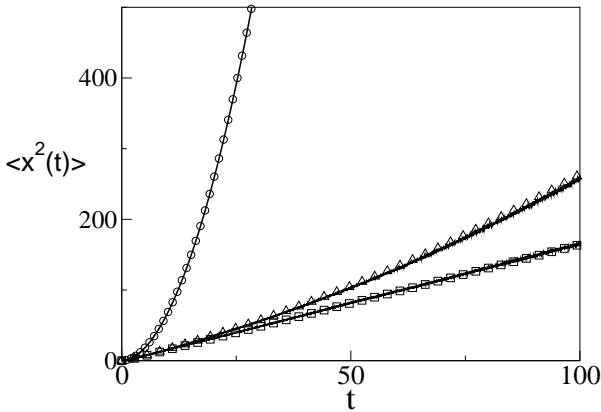}
\caption{Second  moment of the distribution as a function of time
in the case of potential~$U(x)=\varepsilon\frac{x^3}{1+x^2}$.
  Eqs.~(\ref{smp}) and~(\ref{fsmnp}  for an exponential waiting times distribution
$\psi(t)=\frac{\gamma}{2} e^{-\gamma t/2}$  with $\gamma=1$ (triangles and squares) and for a
Mittag-Leffler distribution  with $\alpha=0.5, T =1$  (circles) are plotted and compared to numerical simulations (continuous lines).
The bottom continuous line and square symbols refer to the analytical and numerical case with  $\varepsilon=0.01$ while the other curves correspond 
to the value $\varepsilon=0.1$ used for Fig.1 and 2. Since the potential is not limited, the quadratic correction in Eqs. ~(\ref{smp}) and ~(\ref{fsmnp}) is included in the middle curve (triangles),
while for the bottom curves (smaller $\varepsilon$, squares)  and uppermost curve (circles)
 this correction  is negligible compared to the unperturbed value for the second moment.
}
\vspace{0.2 cm}
 \label{fig2m}
\end{figure}

\section{Numerical simulation of the diffusion process}
The diffusion process can be simulated by implementing numerically
Eq.~(2) and then deriving the corresponding probability
distribution as in the following.  The  dichotomic
fluctuation~$\xi(t)=\pm w$ is obtained by generating the time
intervals between each change of value of the variable~$\xi$
randomly  with the specified waiting times distribution
density~$\psi(\tau)$. Eq.~(2) can be numerically integrated
between two subsequent changes of value of the variable ~$\xi(t)$
and the variable~$x$ accordingly updated. This procedure leads to
a diffusion in the variable~$x$ and average over many realizations
of the fluctuations of~$\xi$ allows one to evaluate the probability
density at each time and position on the real axis. In the case of
a Poissonian process,~$\psi(\tau)=\gamma e^{-\gamma \tau }$, the
comparison of the numerical results with the solution given by
Eq.~(\ref{eqc9_b}) is shown in Fig.~\ref{fig1a}. Fig.~\ref{fig1}
shows the case of non-ergodic fluctuations, i.e. the comparison of
the numerical with the analytical solution provided by
Eq.~(\ref{eqc9_bb}) for a  Mittag-Leffler  distribution of waiting
times~$\psi(\tau)$. In order to produce random numbers distributed
according to  a Mittag-Leffler function with parameters $T$  and
$\alpha$ we adopt the same procedure as described
in~\cite{ca,germano}: we first generate two numbers $n_0, n_1$
uniformly distributed between~$0$ and~$1$ and then convert them
via the following transformation:

\begin{eqnarray}
\tau=-T \log(n_0)
\left[\frac{\sin(\pi \alpha)}{\tan(\pi\alpha n_1)}
-\cos(\pi\alpha)\right]^{\frac{1}{\alpha}},
\end{eqnarray}
the numbers so generated can be proved to be distributed exactly
as a Mittag-Leffler function of parameters~$T$ and~$\alpha$.
Fig.~\ref{fig1}  shows that also in this case the agreement  is
excellent confirming the validity of the approach followed.
Finally we evaluated numerically the first and the second moment
of the distribution, i.e. Eq.~(\ref{fmp}) and Eq.~(\ref{smp}).
Also in this case the agreement between theoretical and numerical
is remarkable, even for the case
of non-limited potential as shown in Figs.~\ref{fig1m}~and~\ref{fig2m}.

\section{Concluding Remarks}
We have studied a stochastic diffusion equation in presence of a
potential~$U(x)$. Considering the case of an odd potential,
$U(-x)=-U(x)$  we showed that a first order approximation solution
can be obtained  by the unperturbed solution  simply replacing the
spatial variable~$x$ with~$x+ U(t)$. We  provided an analytical
solution for the case  of  Poissonian  fluctuations and,
remarkably, for  non-Poissonian fluctuations in the non-ergodic
regime where the time scale of the fluctuations diverges. These
solutions for the probability density of the diffusion are exact
for small perturbation and can be generalized to provide
a theoretical tool that can be employed in many fields of growing interest  where anomalous diffusion emerges in physical \cite{bertolo} and biological systems \cite{strunz2} and compared to other approaches based e.g. on fractional calculus and Continuous Time Random Walk \cite{klafter}.
 Interestingly, in  the Poissonian case,
our solution is an approximate solution of a system of coupled
partial differential equations, Eqs.~(\ref{syst1}), with  coefficient depending on the spatial variable
$x$.
 We supported our conclusions showing that the analytical
solutions are in good agreement with numerical simulations,
confirming that the derived solutions can be applied in all
relevant cases of weakly driven anomalous diffusion produced by
dichotomic fluctuations.

\section{Acknowledgements}M.B. acknowledges financial support
from FONDECYT project no 1110231.

\end{document}